\documentclass[reprint,amsmath,amssymb, aps]{revtex4-1}

\usepackage{graphicx}	
\usepackage{float}
\usepackage{dcolumn}	
\usepackage{bm}		
\usepackage{color}		

\usepackage{lipsum}

\makeatletter
\newcommand*{\balancecolsandclearpage}{%
  \close@column@grid
  \clearpage
  \twocolumngrid
}
\makeatother

\begin{document}

\title{Physics of beer tapping}

\author{Javier Rodr\'{\i}guez-Rodr\'{\i}guez$^1$\email{javier.rodriguez@uc3m.es}, Almudena Casado-Chac\'on$^1$ and Daniel Fuster$^2$\vspace{5mm}}

\affiliation{
$^1$ Fluid Mechanics Group, Carlos III University of Madrid, SPAIN\\
$^2$ CNRS (UMR 7190), Universit\'e Pierre et Marie Curie, Institut Jean le Rond d'Alembert, FRANCE
}


\begin{abstract}
The popular bar prank known in colloquial English as {\em beer tapping}
consists in hitting the top of a beer bottle with a solid object, usually
another bottle, to trigger the foaming over of the former within a few seconds.
Despite the trick being known for long time, to the
best of our knowledge, the phenomenon still lacks scientific explanation.
Although it seems natural to think that shock-induced cavitation enhances the
diffusion of CO$_2$ from the supersaturated bulk liquid into the bubbles by breaking them up, the
subtle mechanism by which this happens remains unknown. Here we show that the
overall foaming-over process can be divided into three stages where different
physical phenomena take place in different time-scales, namely: bubble-collapse
(or cavitation) stage, diffusion-driven stage and buoyancy-driven stage. In the
bubble-collapse stage, the impact generates a train of expansion-compression
waves in the liquid that leads to the fragmentation of pre-existing gas
cavities. Upon bubble fragmentation, the sudden increase of the
interface-area-to-volume ratio enhances mass transfer significantly, which makes the
bubble volume grow by a large factor until CO$_2$ is locally depleted. At that
point buoyancy takes over, making the bubble clouds rise and eventually form buoyant vortex rings
whose volume grows fast due to the feedback between the buoyancy-induced rising
speed and the advection-enhanced CO$_2$ transport from the bulk liquid to the
bubble. The physics behind this explosive process sheds insight into the dynamics of geological phenomena such as limnic eruptions.
\end{abstract}

\maketitle

Understanding the formation of foam in a supersaturated carbonated liquid after an impact on the container involves a careful physical description of a number of processes of great interest in several areas of Physics and Chemistry. In order of appearance in this problem: propagation of strong pressure waves in bubbly liquids, bubble collapse and fragmentation, gas-liquid diffusive mass transfer and the dynamics of bubble-laden plumes and vortex rings. All these phenomena are observed, for instance, in the explosive formation of foam occurring in a beer bottle when it is tapped on its mouth, an effect known as {\it beer tapping}. In this letter, we will use this effect as a convenient system to quantitatively describe the interaction between the processes mentioned above that ultimately leads to the explosive formation of foam that occurs in gas-driven eruptions \cite{ZhangKlingAREPS2006}.
As a consequence of the broad range of phenomena taking part in the overall process, the better understanding of the foam forming process in supersaturated liquids finds application in various fields of natural sciences and technology where similar gas-driven eruptions occur. The dynamics of limnic \cite{ZhangKlingAREPS2006, ZhangNature1996} or explosive volcanic \cite{CashmanSparksGSAB2013, Magan_etalJVGR2004} eruptions and the formation of flavour-releasing aerosols by bursting Champagne bubbles \cite{LigerBelair_etalPNAS2009} are just a few examples. 
Mott and Woods \cite{MottWoodsJVGR2010} have triggered a chain reaction in a stably-stratified tank containing a deep layer of  CO$_2$-saturated lemonade and a shallower layer of fresh water by spilling a gravity current of salt grains along the bottom of the tank. This example shows that the dynamics of CO$_2$ bubbles in daily-life liquids can be used to explain complex natural phenomena such as limnic eruptions.

To understand how the processes described above interact to lead to the foaming-up of beer, we have carried out an experimental investigation impacting commercial beer bottles under well-controlled repeatable conditions (see Supplementary Material). In particular, bubbles have been generated at a fixed location far from the bottle walls by focusing a laser pulse into the bulk liquid. In this way, we avoid the variability in the formation of bubbles upon the impact caused by the arbitrary distribution of nucleation sites \cite{Brennen1995}, thus ensuring that a bubble with a known initial size is always present in the measurement volume. By recording the evolution of these gas bubbles with a high-speed camera and the liquid pressure temporal evolution with a hydrophone  we provide qualitative and quantitative analyses of the various processes that develop during the foam formation. Thus, we divide the overall foaming-over process into well-differentiated stages controlled by different physical mechanisms.  More importantly, we show experimental evidence supporting the explanation given for each step of the outgassing process.

\begin{figure}[h]
\centering
\includegraphics[width=\columnwidth]{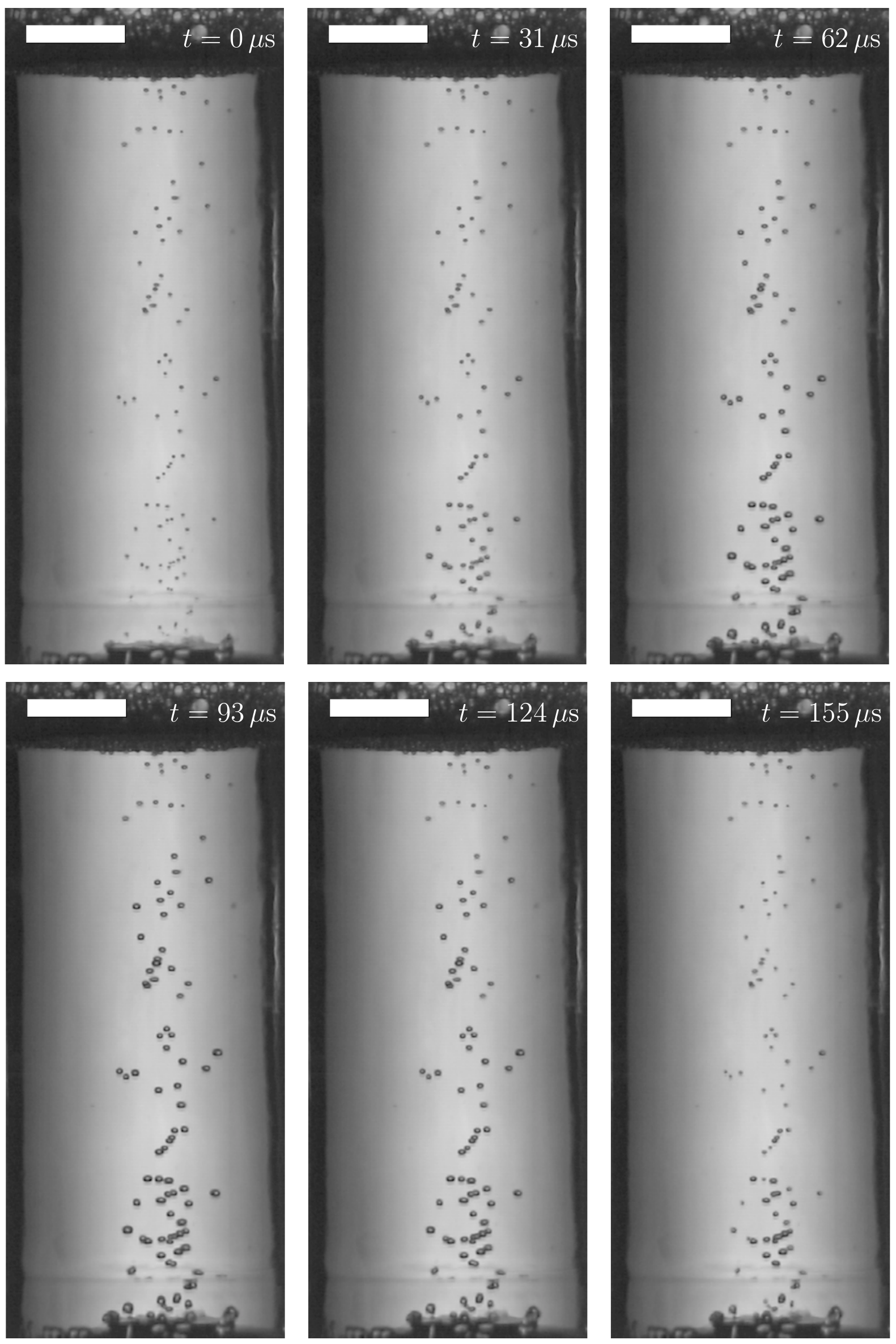}
\caption{\label{fig_pressure_wave}Sequence of images corresponding to the instants right after the pressure wave starts to propagate through the beer. To ensure a continuous bubble cloud, a metal disk was placed at the bottom of the bottle in this experiment, what effectively introduces a large number of nucleation sites. Scale bar: 10 mm.}
\end{figure}

\begin{figure}
\centering
\includegraphics[width=\columnwidth]{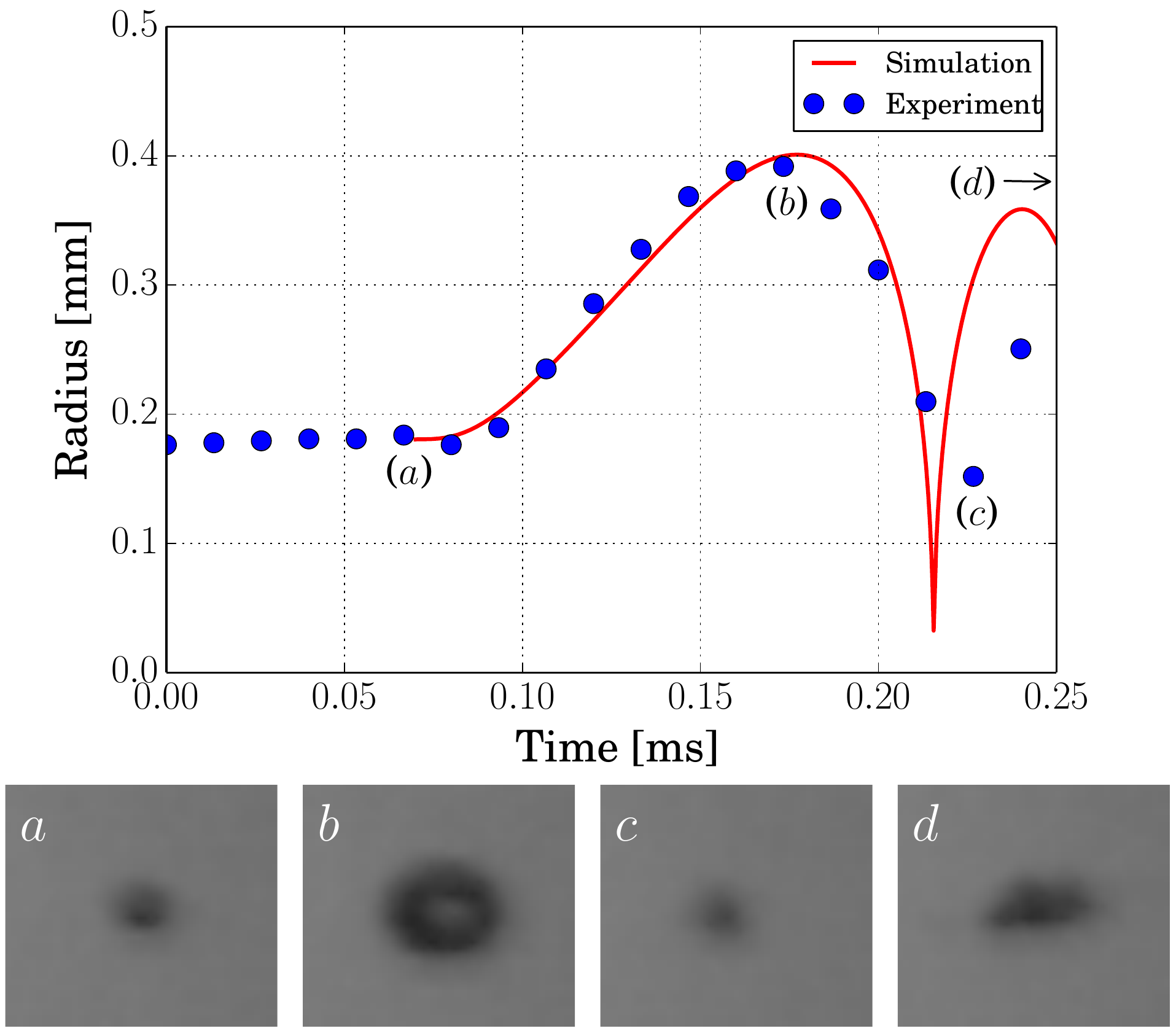}
\caption{\label{fig_bubble_implosion}A bubble of initial radius $R_0 \approx 180$ ${\mu}$m, induced by a laser pulse nearly a second before the impact, grows after the passage of the expansion wave reaching the bubble location at instant ($a$). After the instant of maximum radius ($b$), the bubble collapses at some point between frame ($c$) and the previous one, turning into a bubble cloud. Panel $(d)$ shows the bubble cloud 0.1 ms after instant ($c$). The Rayleigh-Plesset equation has been integrated numerically (red solid line, see Supplementary Material) for a bubble subjected to a pressure pulse with an amplitude, $p_A = 100\,\mathrm{kPa}$, and period, $T = 0.24\,\mathrm{ms}$, measured at the bubble's location with a hydrophone. Notice that, after the implosion, $t \approx 0.22\,\mathrm{ms}$, the Rayleigh-Plesset equation no longer describes the behavior of the bubble since it is only valid for a single bubble, not for a bubble cloud. (see Movies S1 and S2).}
\end{figure}

The chain of events that ultimately leads to the foaming-up of beer is
triggered by a sudden impact on the top of the bottle, which generates a
compression wave that propagates through the glass towards the bottom as
predicted by the classical theory of impact on solids \cite{Goldsmith2001}.
When the wave reaches the base of the bottle, it is partially transmitted to
the liquid as an expansion wave that travels towards the free surface, where it
bounces back as a compression wave. A simple model for
the impact problem \cite{Goldsmith2001} implies that the stiffer the bottle, the more efficient is the
transmission of the expansion wave to the liquid. 
Thus, the shock is more efficiently transmitted into the liquid in the case of a
glass container than in a softer bottle (e.g. plastic), although the expansion wave
is still generated, albeit its smaller amplitude. The train of waves transmitted to the liquid bounce back and forth several times until it damps out. Figure \ref{fig_pressure_wave} shows snapshots of the first instants after the impact to illustrate the effects of the first expansion-compression cycle. It can be seen how bubbles start to expand first near the bottom whereas, at approximately $t \approx 124 \, \mu{\rm s}$, those located near the free surface begin to shrink.  The train of
rarefaction-compression waves drives the fragmentation of most of the existing gas pockets during the first wave cycles.  Figure \ref{fig_bubble_implosion} illustrates a typical time evolution of the bubble radius measured in our experiments along with snapshots showing the bubble at different relevant instants of the expansion-collapse process (see also Movies S1 and S2 in Supplementary Material). This first stage of the overall foaming-up process lasts of the order of the acoustic time of the liquid volume, usually $t_\mathrm{ac} = 2H/c \approx 0.2\,\mathrm{ms}$, assuming a typical liquid height $H \approx 10\,\mathrm{cm}$ and a speed of sound around $c \approx 1000\,\mathrm{m/s}$. Notice that, since waves propagating in bubbly liquids are strongly damped, the intensity of successive rebounded waves decays rapidly thus they are less likely to cause bubble collapse.

Similarly to what happens in the generation of medical ultrasound contrast agents through sonication (ref. \cite{Feinstein_etalJACC1984}) or, albeit in a more violent way, in sonoluminescence \cite{Brenner_etalRMP2002}, it seems reasonable to attribute the break-up of the bubbles to a Rayleigh-Taylor instability \cite{Brenner_etalRMP2002}. The number of fragments, $N$, resulting upon the break up of a bubble cannot be measured due to the high void fraction of the resulting bubble cloud (Fig. \ref{fig_bubble_implosion}d). Instead, an estimation of this number is obtained using the model of Brennen \cite{BrennenJFM2002}, based on the ideas put forward by several authors \cite{ProsperettiSeminaraPoF1978,Ioss_etalPoF1989}. Following this model (see Supplementary Material), the most unstable mode is given by
\begin{equation}
\centering
n_m = \frac{1}{3}\left(\left(7 + 3\Gamma_m\right)^{1/2} - 2\right),
\label{eq:most_unstable_mode}
\end{equation}
with $\Gamma_m = \rho R^2 \ddot{R} / \sigma$, evaluated at the instant when the radius, $R$, is minimum, $\rho$ the fluid density and
$\sigma$ the liquid-gas surface tension. The size of the fragments is
expected to be of the order of $R/n_m$, thus the number of fragments generated is $N \approx n_m^3$. For the
typical bubble sizes and pressure wave amplitudes used in these experiments, we find a
most unstable mode of the order of $n_m \approx 10^2$ and a number of
fragments $N \approx 10^6$.

\begin{figure}[H]
\centering
\includegraphics[width=\columnwidth]{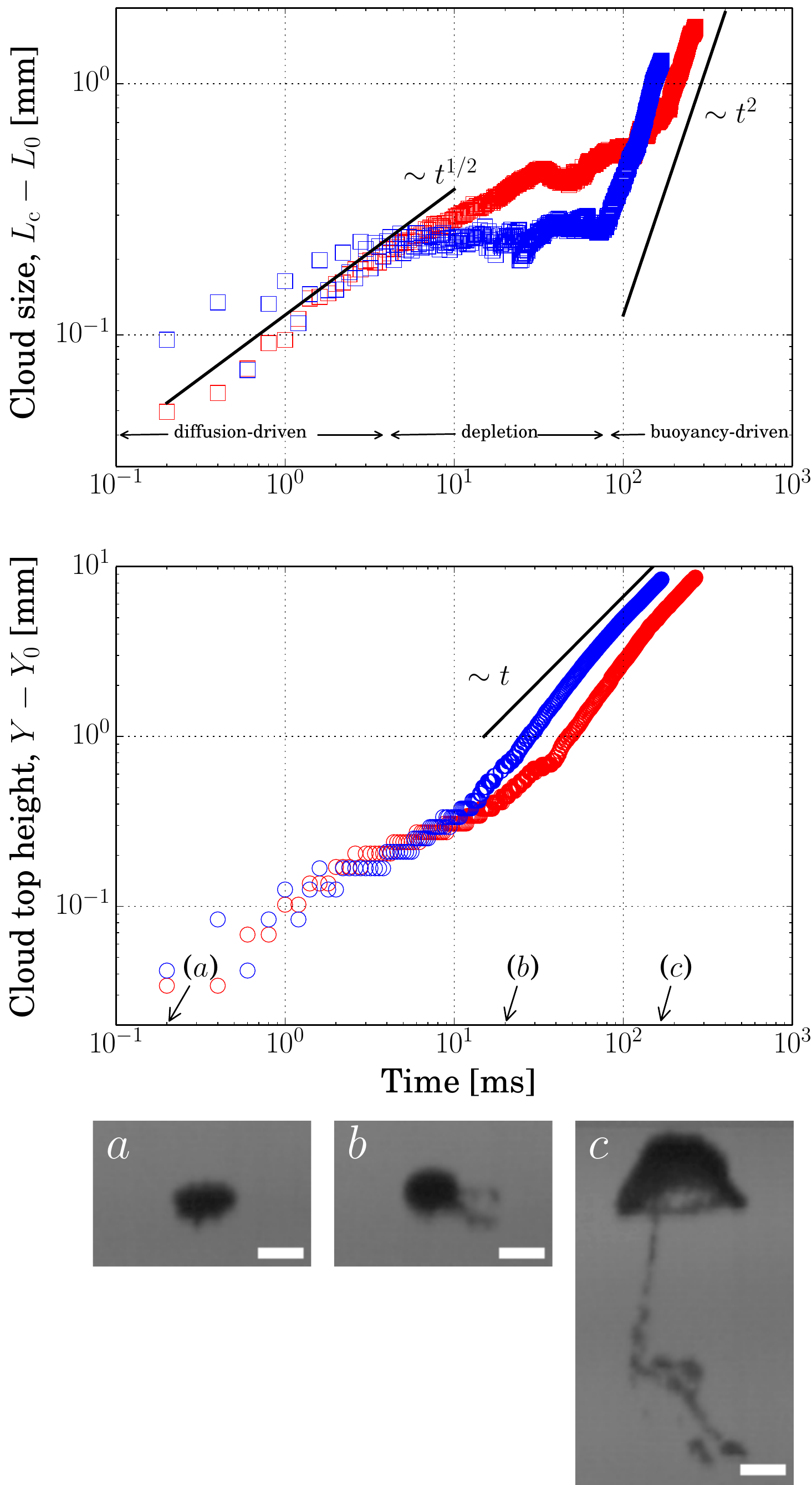}
\caption{\label{fig_cloud_size}{\em Upper plot:} Time evolution of the size of a bubble cloud, $L_\mathrm{c}$, after the shock-induced collapse of an already existing bubble (blue) and after a laser-induced bubble implosion (red). Three different stages have been marked qualitatively: diffusion driven, depletion and buoyancy-driven. Black solid lines depict the scaling laws $L_\mathrm{c} \sim t^{1/2}$ (diffusion-driven) and $L_\mathrm{c} \sim t^2$ (buoyancy-driven). {\em Lower plot}: Time evolution of the top of the bubble cloud corresponding to the same cases of the upper plot. Notice how, at long times, the plume approaches a steady rising velocity. Letters denote the instants corresponding to the images in the lower pannels (scale bar: 1 mm). (See movie S3)}
\end{figure}

As a consequence of the fast bubble collapse and break-up, right after the implosion the total gas-liquid
interfacial area increases by a factor of the order of $N^{1/3}$. This sudden
increase of the interfacial area leads to a second stage where the clouds of
bubble fragments grow rapidly as a result of the diffusion of carbonic
gas into the newly created cavities. This stage can be modelled using the
classical theory of bubble growth in supersaturated
media (Ref. \cite{EpsteinPlessetJCP1950}). Under the reasonable assumption that the
cloud grows as the sum of its components, this theory states that the cloud size, $L_\mathrm{c}$, follows
\begin{equation}
\centering
L_\mathrm{c} = L_0 + \alpha N^{1/3} F\left(\frac{\Delta C}{\rho_\mathrm{g}} \right)\sqrt{\frac{\kappa t}{\pi}},
\end{equation}
where $\Delta C$ is the difference between the concentration of carbonic gas in
the bulk liquid and the saturation value, $\kappa$ its diffusivity,
$\rho_\mathrm{g}$ the density of the gas inside the bubbles, $\alpha$ a dimensionless constant and $F(x)$ a known function (Supplementary Material).
Taking the estimated number of fragments generated during the collapse of a single bubble,
$N \approx 10^6$, we expect the radius of the bubble cloud to grow about 100
times faster than a single bubble with the same volume than the cloud. In fact, this magnitude represents an upper bound, since those fragments at the center of the cloud will grow more slowly, due to their limited access to CO$_2$. Initially, the growth rate scales roughly as $t^{1/2}$  albeit exhibiting some oscillations caused by cycles of expansions and compressions that are not yet attenuated (see the stage labelled as ``diffusion-driven'' in Fig. \ref{fig_cloud_size}, blue squares in the upper panel).
This diffusion-driven stage ends when carbon dioxide is locally depleted and thus the cloud's size significantly moderates its growth.

In the example of figure \ref{fig_cloud_size}, this occurs at about $t \approx 10$ ms (stage labelled as ``depletion''). To avoid the noise introduced by the acoustic waves  at short times we have performed additional experiments by focusing a high-energy laser pulse inside the liquid to trigger the formation of a bubble cloud by laser-induced cavitation \cite{OhlLindauLauterbornPRL1998} (see Supplementary Material). In the focal region, the laser generates a dense bubble cloud that initially grows as the square root of time, which is consistent with a purely diffusive growth (Fig. \ref{fig_cloud_size}, red squares in the upper panel).

The rapidly growing bubble clusters act as buoyancy sources that lead to the formation of bubble-laden buoyant vortex rings in time scales of order $t_g \sim \left(L/g\right)^{1/2}$ (Fig. \ref{fig_cloud_size}c), very much like a localized release of heat forms a thermal \cite{TurnerPRSA1957}. As the vortices rise through the liquid, the advection due to their self-induced velocity and the mixing caused by their vortical motion contribute to enhance the transport of CO$_2$ to the bubbles.
In turn, this results in a growth-rate faster than that
found for pure diffusion, namely $t^{1/2}$. Indeed, the cloud's size grows roughly as
$t^2$ during this stage (Fig. \ref{fig_cloud_size}, upper plot). Moreover, as a
consequence of the continuous generation of gas volume inside the vortex, the
velocity approaches a constant value (Fig. \ref{fig_cloud_size}, lower plot) instead
of decreasing as happens in buoyant vortex rings induced by the release of a
fixed amount of buoyancy. For instance, in thermals originated by the sudden
release of a fixed amount of heat, the velocity decays as $t^{-1/2}$ due to
the entrainment of colder fluid \cite{TurnerPRSA1957,MortonJFM1960}. 

Conversely, in the bubble-laden plumes studied here, the feedback between buoyancy-driven rising motion and gas-volume generation results in a nearly constant speed. 
This behavior is similar to that found in the the so-called autocatalytic vortex
rings or plumes \cite{RogersMorrisPRL2005} where buoyancy is continuously produced by a
chemical reaction that yields products less dense than the reactants. These buoyancy-driven chemically-reacting flows appear, for instance, in the combustion of flame balls in microgravity conditions. Interestingly, they are also relevant in some explosion scenarios for Type Ia supernovae \cite{VladimirovaCTM2007, RogersMorrisChaos2012}.

The analogy between the bubble-laden plumes observed here and the autocatalytic plumes described in the literature extends also to their morphology. Panel $c$ in Fig. \ref{fig_cloud_size} shows one of these bubble-laden plumes when it is well developed (see also Movie S3). The plume consists of a vortex with a nearly spherical cap with a thin conduit that ascends more slowly, features observed in the plumes driven by autocatalytic chemical reactions \cite{RogersMorrisPRL2005, RogersMorrisChaos2012}.

It should be pointed out that, among all the stages of the foaming-up process, this one is the most effective in terms of the amount of liquid outgassing as a result of its self-accelerating nature. This stage starts at times of the order of tens of milliseconds and concludes when the plumes reach the size of the liquid volume, usually of the order of a second.

Remarkably, the behavior of the bubble-laden vortex rings during the diffusion-driven and buoyancy-driven stages is independent of the mechanism used to generate the initial bubble cloud. Figure \ref{fig_cloud_size} depicts the evolution of the bubble cluster size and velocity of a bubble cluster originated by laser-induced cavitation. The size and velocity follow the same scaling laws as the vortex created by the pressure-induced bubble implosion. This suggests that similar explosive CO$_2$ outgassing processes driven by the formation of these bubbly plumes may be initiated by other physical mechanisms generating dense bubble clouds such as the introduction of new bubble nucleation sites \cite{CoffeyAJP2008, WoodsARFM2010} or a sudden change on the saturation conditions occurring either globally \cite{Magan_etalJVGR2004} or locally \cite{ZhangNature1996, WoodsARFM2010}. In fact, our observations suggest that, once the plume is initiated, its dynamics do not seem to depend on the particular initiation mechanism. Thus, one of the main conclusions of this study is that the dynamics of these bubble-laden self-accelerating plumes moving in supersaturated media may partly explain the explosive behavior of systems such as limnic and explosive volcanic eruptions where current models typically neglect the role of these autocatalytic structures \cite{ZhangKlingAREPS2006, WoodsARFM2010, MottWoodsJVGR2010, CashmanSparksGSAB2013, Magan_etalJVGR2004}.

Finally, two side effects induced by the development of the bubbly plumes must
be mentioned here attending to their relevance in the global degassing process in the case of the bottle.
Firstly, due to the finite size of the container, a global recirculating motion
is generated that drags bubbles from near the free surface deep into the bulk
liquid, thus increasing their residence time in the flow and allowing them to
grow for longer times. Secondly, the flow induced inside the bottle speeds-up
also the growth of gas cavities attached to the walls \cite{Jones_etalACIS1999,
Liger-Belair_etalJPCB2005} through the enhancement in the transport of carbon
dioxide towards these cavities, that otherwise would only grow by diffusion. A very similar effect is probably behind the long time scales involved in limnic eruptions. In these phenomena, bubbly plumes form that keep entraining CO$_2$-saturated water from the bottom of the lake until it is almost depleted of this gas due to the global overturning flow that they induce in the lake \cite{MottWoodsJVGR2010}.
Altogether, the chain of effects described in this letter leads to the fast appearance of foam that has granted {\em beer tapping} its popularity.

We would like to thank Dr. Maylis Landeau for her useful comments on the potential applications
of the present study. The authors are indebted to Profs. Norman Riley, Jos\'e Manuel Gordillo, Alejandro Sevilla and Carlos Mart\'{\i}nez-Baz\'an for their insightful comments. We acknowledge the support of the Spanish Ministry of Economy and Competitiveness through grant DPI2011-28356-C03-02.

\balancecolsandclearpage

\section*{Supplementary Material}

{\bf Experimental techniques:}

\begin{figure}[H]
\centering
\includegraphics[width=\columnwidth]{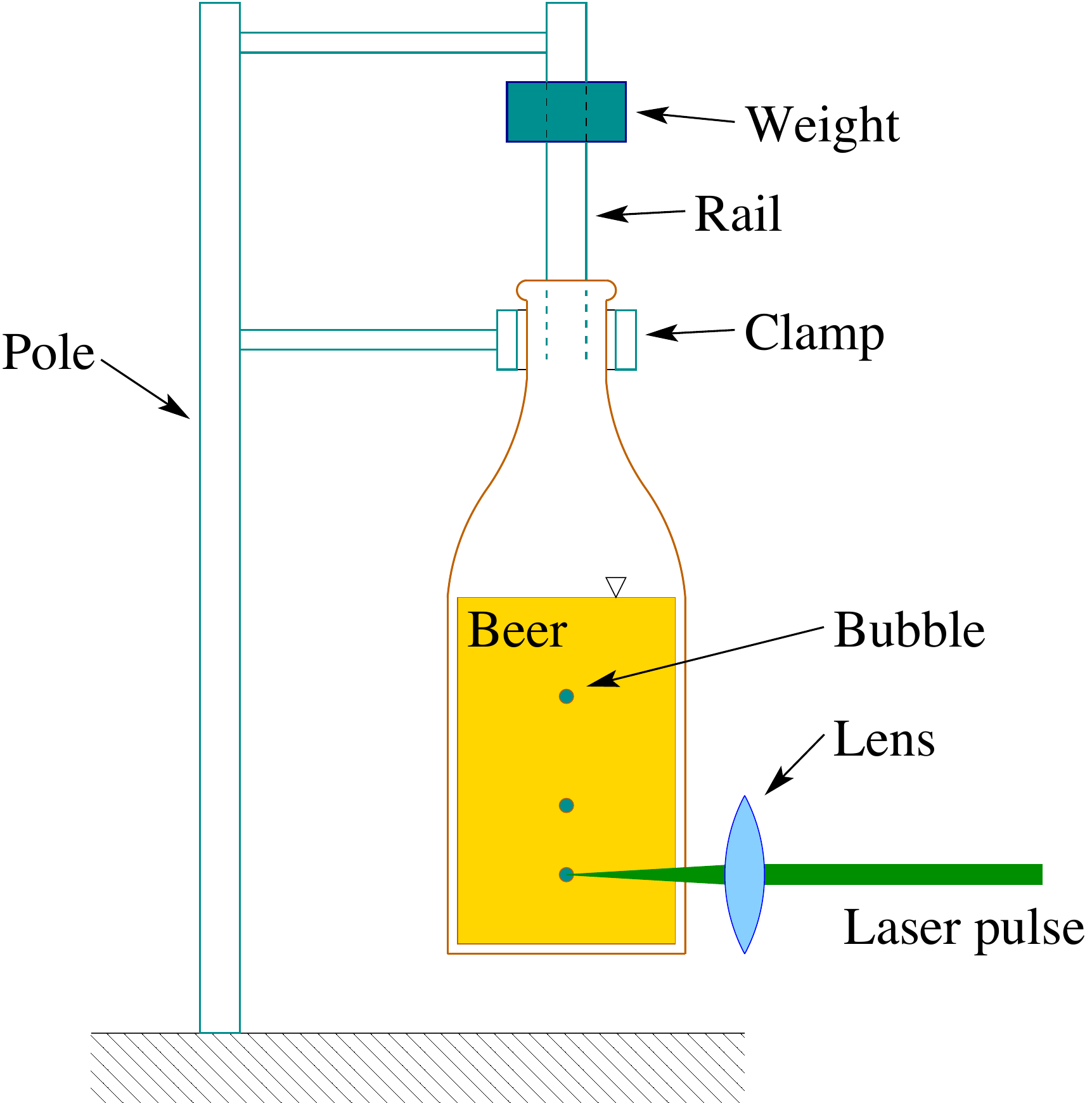}
\caption{\label{fig_experimental_setup}Sketch of the experimental setup. The bottle is held in place with a clamp around its neck that prevents its motion after the impact. To ensure a good repeatability, the weight falls along a rail that guarantees that its lower side remains horizontal during the falling. The YaG laser pulse used to generate both the individual bubbles (low-energy pulse) and the bubble clouds (high-energy pulse) is also depicted.}
\end{figure}

In a first set of experiments, the pressure fluctuations induced in the liquid upon the impact were recorded with a hydrophone. In these experiments, the bottle was filled with deionized water, to delay the formation of cavitation bubbles at the hydrophone's surface.
In a second set of experiments, actual beer was used (CO$_2$ concentration: $5.0 \pm 0.4$ g/l, measurement provided by the manufacturer). The evolution of bubbles existing in the liquid previously to the impact was recorded with a high-speed camera. To ensure that a bubble is present at the camera's focal point, bubbles were generated by focusing a low-energy YaG laser pulse inside the bulk liquid with a convergent lens before the impact. In this way, we avoid the variability in the location and size of the bubble formed upon the impact due to the randomness of the distribution of nucleation sites found in all liquids \cite{Brennen1995}. In these two experimental sets, the bottle was impacted by dropping a brass weight (90 g.) in a controlled and repeatable manner from a height of 25 mm. (see Fig. \ref{fig_experimental_setup}). Moreover, the bottle is held by the neck using a rigid clamp that prevents the net motion of the bottle after the impact and keeps the bottom above the ground, thus preventing the partial transmission of the wave to the bench.

Finally, a third kind of experiment was performed where, instead of hitting the bottle mechanically, a high-energy laser pulse was focused inside the beer to induce cavitation, similarly as described in \cite{OhlLindauLauterbornPRL1998}, thus creating a bubble cloud that serves as a seed to eventually trigger the formation of a bubble-laden plume. In all the sets, the projected area of the bubble cluster, $A$, was measured using custom-made software. The cluster size, $L_\mathrm{c}$, was then calculated as the diameter of a circle with the same area, namely $L_\mathrm{c} = \sqrt{4A/\pi}$.\\

{\bf Estimation of the number of bubble fragments:} Upon bubble collapse, the high void fraction of the resulting cloud of fragments precludes the application of any measurement technique to directly determine the number of those fragments. To get at least an order of magnitude estimation of that number, we apply the model proposed by Brennen \cite{BrennenJFM2002}. Briefly, this model uses the equation for the amplitude, $a(t)$, of a spherical harmonic perturbation of order $n > 1$,
\begin{equation}
\ddot{a} + \frac{3}{R}\dot{R}\,\dot{a} - \left[(n-1)\frac{\ddot{R}}{R} - 
(n-1)(n+1)(n+2)\frac{\sigma}{\rho R^3}\right]a = 0,
\label{eq:harmonic_perturbations}
\end{equation}
where $\rho$ is the liquid density and $\sigma$ the liquid-gas surface tension. Notice that, to determine the evolution of the perturbations, the time-history of the bubble radius, $R(t)$, is needed. As a further simplification, since the growth of the surface instabilities is very fast compared to the bubble radial dynamics, both the radius, $R$, and its acceleration, $\ddot{R}$, are considered constant during the final instants of the break-up process. Thus, it is now possible to calculate the fastest-growing mode, $n_m$, i.e. that maximizes the expression in brackets, $\left[...\right]$, in equation \ref{eq:harmonic_perturbations}:
\begin{equation}
\centering
n_m = \frac{1}{3}\left(\left(7 + 3\Gamma_m\right)^{1/2} - 2\right),
\end{equation}
where $\Gamma_m = \rho R^2 \ddot{R}/\sigma$, evaluated using the minimum bubble radius, $R_\mathrm{min}$ and the acceleration at the time at which that radius is reached. Consequently, Brennen's model allows us to compute the fastest-growing mode using only values of the radius and its acceleration at the collapse time.\\

\noindent Nonetheless, the short duration of the collapse stage impedes recording the process with enough time resolution to estimate either the minimum bubble radius or the acceleration during this stage (see Figs. \ref{fig_bubble_implosion}c and \ref{fig_bubble_implosion}d). Therefore, to obtain these parameters to apply Brennen's model \cite{BrennenJFM2002}, the evolution of the bubble radius has been computed numerically by integrating the Rayleigh-Plesset equation:
\begin{equation}
\begin{split}
\rho R \ddot{R} + \frac{3}{2} \rho \dot{R}^2 - & \left(p_0 + \frac{2\sigma}{R_0}\right)\left(\frac{R_0}{R}\right)^{3\gamma} + \\ + p_0 +\frac{2\sigma}{R} + \frac{4\mu\dot{R}}{R} & = -p_\mathrm{w}(t),
\end{split}
\end{equation}
where $\mu = 10^{-3}$ kg/(m$\cdot$s) is the liquid viscosity, $\gamma=1.304$ is the heat capacity ratio of the gas (CO$_2$) and $p_\mathbf{w}(t)$ is the pressure pulse induced in the system, given by $p_\mathrm{w}(t) = -p_A\sin\left(2\pi t/T\right)$, with amplitude $p_A = 100$ kPa and period $T = 0.24$ ms. These values were measured with a hydrophone at the bubble's location. Moreover, the following values have been used: the CO$_2$-water surface tension, $\sigma = 0.0434$ N/m, the water density, $\rho = 10^3$ kg/m$^3$, and the ambient pressure, $p_0 = 10^5$ Pa.

Figure \ref{fig_bubble_implosion} shows the good agreement between the numerical and the experimental results before the break up, justifying the usage of the numerical parameters to estimate the number of fragments resulting from it. For instance, from the experiment shown in figure \ref{fig_bubble_implosion}, we obtain $n_m \approx 144$. This value ranges between 120-160 in all our experiments, with a maximum different between the maximum radius calculated numerically and the measurement always within 10\% or less.\\

\noindent {\bf Modelling the growth rate of a bubble cloud:} To get an estimation of the growth rate of the bubble cloud due to the diffusion of CO$_2$ before buoyancy becomes important (diffusion-stage), we assume that the cluster's volume grows as the sum of the volume of its components. The time evolution of the radius of a fragment, $R_\mathrm{f}(t)$, is given be the Epstein-Plesset equation (Ref. \cite{EpsteinPlessetJCP1950}),
\begin{equation}
\frac{\mathrm{d}R_\mathrm{f}}{\mathrm{d}t} = \frac{\kappa \Delta C}{\rho_g} \left(\frac{1}{R_\mathrm{f}} + \frac{1}{\sqrt{\pi \kappa t}}\right),
\end{equation}
where $\kappa$ is the diffusivity of CO$_2$ in water, $\rho_g$ the density of the gas inside the bubble and $\Delta C$ the difference between the concentration of CO$_2$ far away from the cloud and that at the bubble's surface. Neglecting the capillary pressure, the solution to the equation is well approximated by
\begin{equation}
R_\mathrm{f}(t) \approx R_\mathrm{f,0} + F\left(\frac{\Delta C}{\rho_g}\right)\sqrt{\frac{\kappa t}{\pi}},
\label{eq:Epstein_Plesset}
\end{equation}
with $R_\mathrm{f,0}$ the initial radius and $F(x) = x\left(1+\sqrt{1+ 2\pi/x}\right)$ (ref. \cite{EpsteinPlessetJCP1950}).
Notice that the second term in this expression corresponds to the solution to Eq. (\ref{eq:Epstein_Plesset}) at long times. Here we have added the first term, the initial radius, to obtain an expression that yields small relative errors for both long and short times. Assuming that the $N$ fragments are equally sized, the volume of the cloud can be written as
\begin{equation}
L_\mathrm{c}^3 = \alpha^3 N \left[R_\mathrm{f,0} + F\left(\frac{\Delta C}{\rho_g}\right)\sqrt{\frac{\kappa t}{\pi}}\, \right]^3,
\end{equation}
where the dimensionless constant $\alpha$ accounts for the fact that the cluster contains some amount of liquid separating the bubbles. Taking the cube root of the expression, and denoting the initial cluster size by $L_0 = \alpha N^{1/3} R_\mathrm{f,0}$, we get
\begin{equation}
L_\mathrm{c} = L_0 + \alpha N^{1/3} F\left(\frac{\Delta C}{\rho_g}\right) \sqrt{\frac{\kappa t}{\pi}}.
\end{equation}
\noindent Finally, we would like to comment on the applicability of the Epstein-Plesset equation to this problem. This equation describes the growth of an isolated bubble, whereas in this case bubbles belong to a dense cluster. Therefore we expect that, at some point, the CO$_2$ must be depleted, with only those bubbles closer to the clouds' surface being still able to grow. This is consistent with the behavior observed in figure \ref{fig_cloud_size}b, where the growth rate of the cluster decreases notably at about 10 ms.\\

\section*{Movie labels:}

{\bf Movie S1. Bubble fragmentation 1:} fragmentation of a bubble as a result of the passage of the train of expansion-compression waves. This is the same bubble shown in Fig. 2. Scale bar: 500 $\mu$m.\\

\noindent {\bf Movie S2. Bubble fragmentation 2:} the fragmentation of a gas bubble is shown, together with a plot of its instantaneous radius. Notice how, after the second compression, the bubble breaks-up and forms a cloud of bubble fragments.\\

\noindent {\bf Movie S3. Bubble plume development:} this movie shows the evolution of a bubble cloud during the buoyancy-driven stage. Notice the large increase of the foam volume. This plume corresponds to the blue symbols of Fig. 3. Scale bar: 1 mm.\\

\noindent {\bf Movie S4. Global view of the bottle after the impact:} the movie shows the development and growth of the bubble plumes after hitting the bottle. To make more evident the process, the formation of bubbles has been forced by placing a metal disk at the bottom of the bottle, what effectively introduces a large number of nucleation sites at the bottom.

\end{document}